\documentclass[journal]{IEEEtran}
\usepackage{amsmath,amsfonts}
\usepackage{algpseudocode}
\usepackage{algorithm}
\usepackage{array}
\usepackage[caption=false,font=normalsize,labelfont=sf,textfont=sf]{subfig}
\usepackage{textcomp}
\usepackage{stfloats}
\usepackage{url}
\usepackage{verbatim}
\usepackage{graphicx}
\usepackage{cite}

\begin{document}

\title{Optimizing Multi-Lane Intersection Performance in Mixed Autonomy Environments}

\author{Manonmani Sekar and Nasim Nezamoddini%
\thanks{Industrial and Systems Engineering Department, Oakland University, MI, USA. 
E-mail: nezamoddini@oakland.edu.}}


\maketitle

\begin{abstract}
Abstract—One of the main challenges in managing traffic at multilane intersections is ensuring smooth coordination between
 human-driven vehicles (HDVs) and connected autonomous vehicles (CAVs). This paper presents a novel traffic signal control framework that combines Graph Attention Networks (GAT) with Soft Actor-Critic (SAC) reinforcement learning to address this challenge. GATs are used to model the dynamic graph- structured nature of traffic flow to capture spatial and temporal dependencies between lanes and signal phases. The proposed SAC is a robust off-policy reinforcement learning algorithm that enables adaptive signal control through entropy-optimized decision making. This design allows the system to coordinate the signal timing and vehicle movement simultaneously with objectives focused on minimizing travel time, enhancing performance, ensuring safety, and improving fairness between HDVs and CAVs. The model is evaluated using a SUMO-based simulation of a four-way intersection and incorporating different traffic densities and CAV penetration rates. The experimental results demonstrate the effectiveness of the GAT-SAC approach by achieving a 24.1\% reduction in average delay and up to 29.2\% fewer traffic violations compared to traditional methods. Additionally, the fairness ratio between HDVs and CAVs improved to 1.59, indicating more equitable treatment across vehicle types. These findings suggest that the GAT-SAC framework holds significant promise for real-world deployment in mixed-autonomy traffic systems.
\end{abstract}


\section{Introduction}
\IEEEPARstart{T}{he} Urban traffic congestion has emerged as one of the most pressing challenges facing modern cities, with significant implications for economic productivity, environmental sustainability, and quality of life. Traditional traffic control systems, which predominantly rely on fixed-time signal plans~\cite{Webster1958Traffic} or rule-based adaptive methods such as SCOOT~\cite{Hunt1981SCOOTA} and SCATS~\cite{Lowrie1982SCATS,Sims1979The}, have demonstrated limited capacity to accommodate the dynamic and stochastic nature of contemporary traffic demands~\cite{Papageorgiou2003Review}. These conventional approaches were designed for relatively homogeneous traffic streams consisting primarily of human-driven vehicles, and their performance degrades substantially when faced with the complexities introduced by mixed autonomy scenarios.

The advent of connected and automated vehicles (CAVs) represents a transformative development in transportation systems, offering unprecedented opportunities for improved coordination, efficiency, and safety~\cite{Shladover2018Coexistence,Talebpour2016Influence}. CAVs possess several advantageous characteristics compared to conventional human-driven vehicles (HDVs): they exhibit deterministic behavior, maintain precise vehicle control, respond to control signals with minimal latency (on the order of milliseconds), and can communicate with infrastructure and other vehicles through vehicle-to-everything (V2X) technologies~\cite{Tong2019Artificial,Li2020Intersection}. These capabilities enable cooperative driving behaviors, optimized trajectory planning, and real-time adaptation to traffic conditions that were previously unattainable with human drivers alone.

However, the transition to fully automated transportation will necessarily involve an extended period during which CAVs and HDVs must coexist on shared infrastructure. This mixed autonomy environment introduces substantial modeling and control challenges that cannot be adequately addressed by existing traffic management systems. Human drivers exhibit significantly different operational characteristics compared to automated vehicles: reaction times typically range from 0.7 to 2.0 seconds~\cite{Olson1986PerceptionResp}, driving behaviors vary considerably across individuals and contexts~\cite{Farah2007Study}, and compliance with traffic regulations and control directives is neither guaranteed nor uniform. Research has demonstrated that even modest fractions of HDVs can disrupt the coordinated behavior of CAV platoons, leading to traffic instability, increased delays, and elevated collision risks~\cite{Olia2019MixedTraffic,Li2020HeterogeneousModel}.

The heterogeneous nature of mixed traffic demands sophisticated modeling frameworks capable of capturing the diverse behavioral characteristics of different vehicle types. Microscopic traffic flow models such as the Intelligent Driver Model (IDM)~\cite{Treiber2000Congested} and lane-changing models like MOBIL~\cite{Kesting2007General} provide foundations for representing individual vehicle dynamics, but must be extended to accommodate the distinct response patterns and decision-making processes of automated systems. Continuum models that incorporate mixed autonomy~\cite{Li2020HeterogeneousModel} have revealed complex interactions between vehicle classes, demonstrating that the benefits of automation can be either amplified or diminished depending on penetration rates, spatial distributions, and control strategies employed.

A critical consideration in designing traffic control systems for mixed autonomy environments concerns fairness and equity across different vehicle classes and user groups~\cite{Zhang2020Fairness,Chen2021EquityAware,Litman2015Evaluating}. Optimization strategies that exclusively prioritize overall system efficiency may inadvertently create substantial disparities in service quality, systematically disadvantaging certain user groups—typically human drivers—while disproportionately benefiting others. Such inequitable outcomes are ethically problematic and may also hinder public acceptance of automated vehicle technologies. Transportation equity research~\cite{Litman2015Evaluating} emphasizes the importance of distributional impacts in infrastructure design and control policies, suggesting that fairness considerations should be explicitly incorporated into optimization objectives rather than treated as afterthoughts.

Safety represents another paramount concern in mixed traffic operations. The introduction of automated vehicles with different sensing capabilities, decision-making algorithms, and motion dynamics alters fundamental assumptions underlying traditional safety analysis~\cite{Archer2005Safety,Porter2013A}. Surrogate safety measures such as time-to-collision (TTC), post-encroachment time (PET), and deceleration rates must be adapted to account for the heterogeneous nature of vehicle interactions~\cite{Hayward1972Nearmiss}. Network-level safety assessment frameworks~\cite{Chen2023NetworkLevel} are required to evaluate system-wide implications of control policies, particularly at signalized intersections where conflicts between different traffic streams are most pronounced.

Recent advances in machine learning, particularly reinforcement learning (RL) and multi-agent reinforcement learning (MARL), have demonstrated considerable promise for adaptive traffic control~\cite{Fan2023Research,Wang2022Multiagent,Ahmadi2021Smart}. Early applications of RL to traffic signal control~\cite{Tan1993Multiagent,Thorpe1996Vehicle} established the feasibility of learning-based approaches, while subsequent developments incorporating deep neural networks~\cite{Li2016Traffic} and policy gradient methods~\cite{Schulman2017Proximal,Mnih2016Asynchronous} have substantially improved performance and scalability. The Soft Actor-Critic (SAC) algorithm~\cite{Haarnoja2018Soft} represents a state-of-the-art approach for continuous control problems, offering stable learning through entropy regularization and off-policy updates. Applications of SAC to traffic signal control~\cite{Wei2019An} have demonstrated superior sample efficiency and robustness compared to alternative methods.

The representation of traffic systems as graphs provides a natural framework for modeling spatial relationships and interactions among vehicles, lanes, and intersections~\cite{Chen2021Graph,Gao2024A}. Graph neural networks (GNNs), particularly those employing attention mechanisms~\cite{Veličković2018Graph}, enable adaptive information aggregation that respects the underlying topology of transportation networks. Graph convolutional approaches have been successfully applied to traffic forecasting~\cite{Li2017Diffusion,Yu2019Spatiotemporal} and traffic signal control~\cite{Chen2021Urban}, demonstrating the capability to capture complex spatial-temporal dependencies. Recent work has explored attention-based architectures for traffic speed prediction~\cite{Ham2023Rethinking} and long-term flow forecasting~\cite{Song2024Graph}, achieving state-of-the-art results through adaptive feature weighting and multi-scale temporal modeling.

Despite these advances, existing approaches to traffic control in mixed autonomy environments exhibit several important limitations. First, most proposed methods do not explicitly differentiate between CAV and HDV agents in their state representations or learning processes, treating all vehicles as homogeneous entities~\cite{Li2020HeterogeneousModel}. This simplification precludes the system from exploiting the distinct characteristics of different vehicle types or adapting control strategies to varying penetration rates. Second, fairness considerations are typically absent from objective formulations, potentially leading to solutions that optimize aggregate metrics while creating substantial inequities~\cite{Zhang2021Quantifying}. Third, the selection of hyperparameters for learning-based controllers often relies on manual tuning or limited grid search, introducing subjectivity and potentially compromising generalization across different scenarios.

This work addresses these limitations through the development of a Graph Attention Network–Soft Actor-Critic (GAT-SAC) framework specifically designed for traffic signal control in mixed autonomy environments. The proposed architecture integrates graph attention networks for spatial reasoning with the SAC reinforcement learning algorithm for adaptive policy learning, while explicitly modeling the behavioral heterogeneity between CAVs and HDVs. The framework incorporates fairness-aware reward formulations that balance efficiency and equity objectives, and employs Tree-structured Parzen Estimator (TPE) optimization for automated hyperparameter tuning across diverse traffic scenarios.

The primary contributions of this research are:

\begin{itemize}
    \item Integrated GAT-SAC Architecture: A novel framework combining spatial reasoning (via GAT) and adaptive control (via SAC) for mixed-traffic signal optimization.
    \item Fairness-Aware Control:Explicit consideration of HDV-CAV behavioral variations is required to achieve equal traffic performance and safety.
    \item {Automated Optimization and Evaluation:} A three-phase experimental methodology employing TPE-based hyperparameter tuning, comparative scenario analysis, and production-scale training.
   \end{itemize}

The remainder of this paper is organized as follows. Section 2 reviews related work in traffic signal control, reinforcement learning applications, graph neural networks, and fairness considerations in transportation systems. Section 3 presents the problem formulation, including the formal definition of the mixed autonomy traffic control problem and the multi-objective optimization framework. Section 4 describes the proposed GAT-SAC architecture in detail, covering the graph attention mechanism, the SAC learning algorithm, and the integration strategy. Section 5 discusses the experimental design, including traffic scenarios, simulation configuration, baseline methods, and evaluation metrics. Section 6 presents results and analysis, examining performance across different CAV penetration rates, traffic demands, and fairness-efficiency trade-offs. Section 7 concludes with a discussion of findings, limitations, and directions for future research.

\section{Related Work}

Managing mixed autonomy traffic at intersections requires understanding several interconnected research areas: adaptive traffic signal control, reinforcement learning, graph neural networks, and fairness-aware optimization. This section examines how these fields have evolved and identifies the gaps that motivate the current work. Traffic signal control has progressed through distinct phases over the past several decades. The earliest systems used fixed timing plans based on Webster's formula~\cite{Webster1958Traffic}, which optimized signal cycles for average historical traffic patterns. While simple to implement, these systems could not respond to real-time changes in traffic conditions. The second generation brought adaptive control systems like SCOOT~\cite{Hunt1981SCOOTA} and SCATS~\cite{Lowrie1982SCATS,Sims1979The}, which adjusted timing based on detector data. These systems represented a significant improvement in responsiveness, but their reliance on hand-crafted rules and simplified traffic models limits their effectiveness in complex, mixed autonomy scenarios~\cite{Papageorgiou2003Review}. More recently, third-generation approaches have emerged using optimization and data-driven learning. Model Predictive Control frameworks can optimize future signal phases but become computationally expensive for large networks. Recent reviews~\cite{Fan2023Research,Chen2024Research} have highlighted the growing interest in machine learning methods that can learn control policies directly from data without requiring explicit traffic models.

Reinforcement learning is one of the techniques that has gained traction as a flexible approach to adaptive traffic signal control. The core idea is to let agents learn effective policies by interacting with the environment and receiving feedback. Early work applied tabular methods like Q-learning and SARSA~\cite{Tan1993Multiagent,Thorpe1996Vehicle} to single intersections. As computational power increased, deep learning enabled more sophisticated approaches such as Deep Q-Networks (DQN)~\cite{Li2016Traffic} and Prximal Policy Optimization(PPO)~\cite{Schulman2017Proximal} to handle high-dimensional state spaces and continuous action domains. The Soft Actor-Critic algorithm~\cite{Haarnoja2018Soft} represents an important advance in continuous control through maximum entropy reinforcement learning. Traditional RL methods focus solely on maximizing expected rewards, which can lead to overly deterministic policies with poor exploration. To solve this issue, SAC takes a different approach by optimizing an entropy-regularized objective that encourages some randomness in the policy. This strategy maintains exploration throughout training and avoids getting stuck in local optima. Empirical studies have shown that SAC offers better sample efficiency and more stable convergence compared to methods such as PPO~\cite{Schulman2017Proximal} and A3C~\cite{Mnih2016Asynchronous}. Given the exploration power of the SAC in handling high dimentional spaces, it was applied for traffic signal control problems~\cite{Wei2019An} and demonstrated faster convergence compared to DQN-based approaches. In traffic management problems, coordinating multiple intersections requires multi-agent reinforcement learning (MARL). In the related literature for MARL applications, some approaches treat each intersection as an independent learner~\cite{Tan1993Multiagent}, while others use centralized training with decentralized execution or counterfactual methods like COMA~\cite{Foerster2018COMA} to improve coordination through shared learning. However, most MARL-based traffic controllers treat intersections or lanes as isolated units without explicitly modeling the spatial relationships that are crucial to understanding traffic flow~\cite{Chen2021Graph,Chen2021Urban,Wei2021Hierarchical}.

One of the efficient techniques to model such relations is using GNN that have fundamentally changed how researchers model transportation systems. By representing roads, intersections, and vehicles as graphs, GNNs can capture structural relationships that traditional neural networks miss. Graph Convolutional Networks~\cite{Defferrard2016Convolutional,Li2017Diffusion} propagate information through connected nodes to learn spatial patterns, while Graph Attention Networks~\cite{Veličković2018Graph,Xu2018How} go further by learning which connections matter most in different contexts. GNNs have achieved impressive results in traffic forecasting~\cite{Ham2023Rethinking,Song2024Graph,Jyothi2024A}, flow prediction~\cite{Li2017Diffusion,Yu2019Spatiotemporal}, and trajectory modeling~\cite{Wang2023Lane}. Some recent work has combined GNNs with reinforcement learning for multi-intersection control~\cite{Chen2021Urban}. However, a common limitation is that these models typically assume all vehicles behave similarly, ignoring the substantial differences between automated vehicles and human drivers. 

This assumption contradicts with current and future's modern transportation systems in which 
    connected autonomous vehicles and human-driven vehicles share the roads~\cite{Shladover2018Coexistence}. Research has shown that even modest levels of CAV penetration can improve traffic stability and throughput when vehicles coordinate effectively~\cite{Talebpour2016Influence}. However, most analyses assume vehicles behave uniformly, which does not reflect reality. The differences between vehicle types are substantial: CAVs can react in 10–100 milliseconds~\cite{Olson1986PerceptionResp}, while human reaction times typically range from 0.7 to 2 seconds~\cite{Board2022Highway}. This timing asymmetry creates coordination challenges and safety concerns. Studies of mixed traffic flow~\cite{Li2020HeterogeneousModel,Olia2019MixedTraffic} have revealed that the benefits of automation depend heavily on how the system accounts for behavioral diversity. Path planning and car-following models designed for uniform fleets often perform poorly when vehicle types are mixed~\cite{Jiang2022Integrated,Tian2024Path}. This underscores the importance of adaptive control strategies that explicitly recognize and respond to behavioral heterogeneity. In addition to heterogeneity considerations, the controller should consider fairness between different types of vehicles. In majority of the RL-based controllers the efficiency-focused
reward functions ignore fairness considerations by over-prioritizing certain actions at the expense of others ~\cite{Zhang2021Quantifying}. On the other hand, including fairness constraints reduces the overall efficiency of the systems~\cite{Shirasaka2023Distributed}. In mixed autonomy settings, fair treatment of both HDVs and CAVs is particularly important for social acceptance and safety~\cite{Zhang2020Fairness,Chen2021EquityAware}. If automated vehicles receive systematically better service, human drivers may perceive the system as unfair, potentially affecting public support for automation. 

In summary, the review of the existing literature highlights the following research gaps:







\begin{itemize}
    \item Most work addresses traffic signal timing and vehicle-level coordination separately rather than treating them as interconnected problems requiring holistic solutions.
    
    \item Current traffic control models typically assume uniform vehicle dynamics, failing to account for the substantial differences in reaction times, decision-making patterns, and capabilities between HDVs and CAVs and decision fairness for their traffic control.
    
    \item The combination of structural awareness and spatial relations between intersections and continuous, stable control has not been fully explored for traffic management applications.
\end{itemize}

These limitations motivate the proposed framework, which integrates Graph Attention Networks with Soft Actor-Critic to achieve adaptive, interpretable, and equitable control in mixed autonomy traffic environments. By explicitly modeling vehicle heterogeneity and incorporating fairness considerations into the optimization objective, this work aims to address the identified gaps and advance the state of the art in intersection control for mixed autonomy scenarios.

\section{Problem Formulation}
\label{sec:problem}

We consider a signalized intersection operating under mixed autonomy traffic, where Connected Autonomous Vehicles (CAVs) and Human-Driven Vehicles (HDVs) coexist. Let the vehicle set at time $t$ be 
$\mathcal{V}(t) = \mathcal{V}_{\text{CAV}}(t) \cup \mathcal{V}_{\text{HDV}}(t)$, and the CAV penetration rate be 
$\rho_{\text{CAV}}(t) = \frac{|\mathcal{V}_{\text{CAV}}(t)|}{|\mathcal{V}(t)|}$. 
CAVs exhibit an average reaction time of 0.1~s, while HDVs respond with an average delay of 1.5~s.

Traffic management includes a different set of decisions for lane channelization, flow allocations, and signal timing control. Lane channelization is a fundamental design principle in intersection management, in which each lane is designated for specific movements—typically left turn, movement, or right turn—to minimize conflict in the context of mixed autonomy traffic systems comprising both CAVs and HDVs.
Traditional static channelization assumes uniform driver behavior and fixed flow patterns. HDVs may require more conservative separation due to variability in reaction time and maneuvering behavior. Once lanes are classified, we allocate the vehicle flows across them. Flow allocation determines the optimal number of vehicles that transition between lanes. In parallel with flow control, we also need to determines the signal timing parameters for each control cycle. The action vector $a_t$ includes:

\begin{itemize}
    \item $G_k$: Green time duration for phase $k$
    \item $\Delta_k$: Phase switching decision
    \item $C_k$: Clearance time between phases
\end{itemize}

 Signal timing optimization is constrained by standard traffic control bounds for minimum and maximum durations for phase duration and green time:

\begin{align*}
T_{\min} \leq {\Delta_k} &\leq T_{\max} \\
G_{\min} \leq G_k &\leq G_{\max} \\
C_k &\geq C_{\min}
\end{align*}

We are looking for the best traffic control strategies that jointly optimizes traffic efficiency, safety, and fairness within the mixed-autonomy environment. 
At each simulation step $t$, the agent evaluates three primary objectives:

\begin{itemize}
    \item {Delay Cost} $D(t)$: Measures the cumulative waiting time of all vehicles in the system, serving as an indicator of traffic efficiency.
    \begin{equation}
D(t) = \frac{1}{|\mathcal{V}(t)|} \sum_{i \in \mathcal{V}(t)} \max(0, t - t_i^{\text{arrival}} - t_i^{\text{freeflow}})
\end{equation}
    \item {Fairness Cost} $F(t)$: Represents the disparity between the average waiting times of Human-Driven Vehicles (${d}_{\text{HDV}}$) and Connected and Automated Vehicles (${d}_{\text{CAV}}$), encouraging equitable signal control. This ensures equitable performance between CAVs and HDVs \cite{Zhang2020Fairness,Chen2021Equity}.
    \begin{equation}
F(t) = \frac{|\bar{d}_{\text{HDV}}(t) - \bar{d}_{\text{CAV}}(t)|}{\max(\bar{d}_{\text{HDV}}(t), \bar{d}_{\text{CAV}}(t))}
\end{equation}
    \item {Safety Cost} $S(t)$: The third term introduces a soft penalty for safety-related violations by relaxing hard safety constraints into differentiable penalty functions to enable gradient-based optimization.
    \begin{equation}
S(t) = \alpha\,\text{RLR}(t) + \beta\,\text{TTC}(t) + \delta\,\text{HB}(t)
\end{equation}
where RLR, TTC, and HB represent red-light violations, 
time-to-collision conflicts, 
and hard-braking events. 
\end{itemize}

 The total multi-objective cost function is expressed as:
\begin{equation}
C_{total}(t) = w_d D(t) + w_f F(t) + w_s S(t)
\label{eq:multiobj}
\end{equation}
where $w_d$, $w_f$, and $w_s$ denote the respective weighting coefficients for delay, fairness, and safety objectives. 

This formulation captures the key challenges of mixed autonomy intersections—behavioral heterogeneity, safety-risk trade-offs, and multi-objective optimization under uncertainty.

\section{Methodology}
\label{sec:methodology}

\begin{figure}[!ht]
\centering
\includegraphics[width=0.45\textwidth]{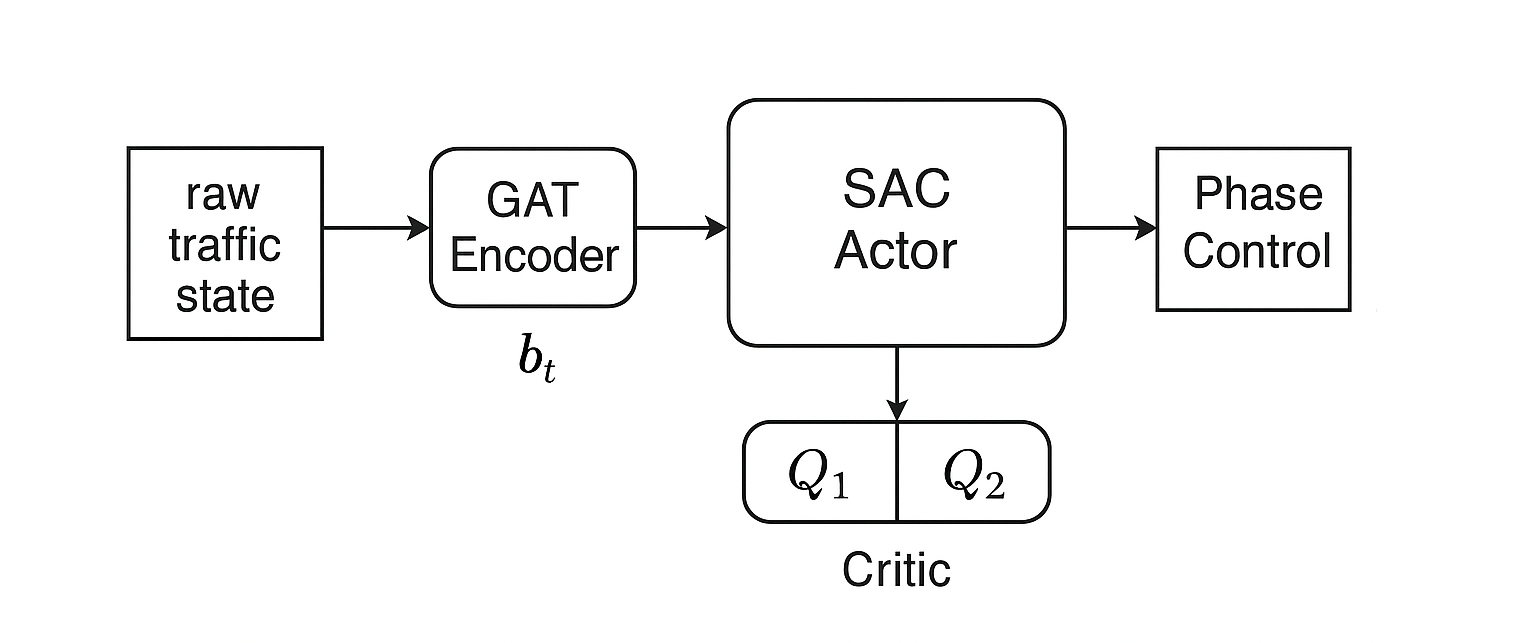}
\caption{Graph Attention Network–Soft Actor–Critic framework }
\label{fig:gat_sac}
\end{figure}
This study proposes a hybrid {Graph Attention Network–Soft Actor–Critic (GAT–SAC)} framework as shown in  Figure \ref{fig:gat_sac} for adaptive traffic signal control in mixed autonomy environments containing both CAVs and HDVs. 
The methodology integrates graph-based spatial reasoning \cite{Veličković2018Graph}, entropy-regularized reinforcement learning \cite{Haarnoja2018Soft}, and multi-objective optimization to enable scalable, fair, and safe traffic management. The framework comprises three principal components:
\begin{enumerate}
    \item \textit{Traffic and environment modeling} representing traffic flow states and its environment setting with intersections, lanes, and flow dependencies.
    \item \textit{Graph Attention Network (GAT)} encoder that models spatial correlations among intersections and lanes.
    \item \textit{Soft Actor–Critic (SAC)} reinforcement learning agent that optimizes lane flow allocation and signal timing policies.
\end{enumerate}


\subsection{Traffic and Environment Modeling}
Traffic signal control inherently operates under conditions of partial observability. The complete state of the traffic network—including precise vehicle positions, speeds, accelerations, origins, destinations, and driver intentions—remains hidden from direct observation. The Partially Observable Markov Decision Process (POMDP) framework provides the theoretical foundation for decision-making under such uncertainty. 
The problem is modeled as a POMDP defined by the tuple $(S, A, O, T, Z, R, \gamma)$, where

\begin{itemize}
    \item $S$: State space
    \item $A$: Action space
    \item $O$: Observation space
    \item $T: S \times A \times S \rightarrow [0,1]$: Transition function
    \item $Z: S \times A \times O \rightarrow [0,1]$: Observation function
    \item $R: S \times A \rightarrow \mathbb{R}$: Reward function
    \item $\gamma \in [0,1)$: Discount factor
\end{itemize}

State space is represented by $S_t = \{x_i^t \in \mathbb{R}^d \mid i = 1, 2, \ldots, N\}$ where $x_i$ denotes the feature vector for intersection or lane that $x_i = [\bar{v}_i/v_0, \ \rho_i, \ \gamma_i, \ q_i]$
 with $\bar{v}_i$ as mean speed, $\rho_i$ as density, $\gamma_i$ as CAV ratio, and $q_i$ as queue length \cite{li2023graph}.  
This representation captures both macroscopic and microscopic traffic characteristics. The action space is $a_t = \{ a^{lane}, a^{signal},a^{conflict}\}$ where $a^{lane}$ is the decision on lane change and flow control weights conditioned on the global intersection state $s_t$, $a^{signal}$ is the decision on signal change and phase duration, and $a^{conflict}$ representing conflicts resolution decisions. Observation Space is $o_t = \{c_{1,t}, c_{2,t}, \ldots, c_{M,t}, \phi_t\} $ where $c_{m,t}$ the reading from vehicle $m$ at time $t$. The transition function is
$T(s_{t+1}|s_t, a_t) = P(s_{t+1}|s_t, a_t)$. The observation function is $Z(o_t|s_t, a_t) = P(o_t|s_t, a_t)$ and the belief state representation is given by $b_t(s) = P(s_t = s | o_1, a_1, \ldots, o_t, a_{t-1})$ and the belief next state is 
$b_{t+1}(s') = \eta \cdot Z(o_{t+1}|s', a_t) \sum_{s \in S} T(s'|s, a_t)b_t(s)$.

A custom simulation environment was developed using the \texttt{Gymnasium} and SUMO (Simulation of Urban Mobility) version 1.15.0. 
The traffic network is modeled as a directed graph and 
vehicle motion follows the Intelligent Driver Model (IDM) 
to simulate realistic longitudinal dynamics. IDM allows for simulating a reasonable number of vehicles and episodes without excessive
computational load, while still providing a more realistic foundation for vehicle movement than simple kinematic models. It can also serve as a basis for modeling HDV behavior. CAVs are modeled with reduced reaction times and smoother acceleration control, whereas HDVs exhibit stochastic variability to reflect human driving behavior.





\subsection{Graph Attention Network (GAT)}
\label{subsec:gat}

To capture spatial dependencies among vehicles and lanes, the intersection state is represented as a directed graph $\mathcal{G}_t = (\mathcal{V}_t, \mathcal{E}_t)$, where nodes correspond to vehicles or lane segments, and edges represent car-following or crossing interactions and vehicle transitions. Each node $v_i$ is described by the normalized state vector $x_i$.
Given input features $x_i$ and $x_j$, the unnormalized attention coefficient is computed as:
\begin{equation}
e_{ij} = \text{LeakyReLU}\left(a^\top [W x_i \Vert W x_j]\right)
\end{equation}
where $W$ is a learnable transformation and $a$ an attention vector.  
Coefficients are normalized via softmax:
\begin{equation}
b_{ij} = \frac{\exp(e_{ij})}{\sum_{k \in \mathcal{N}(i)} \exp(e_{ik})}
\end{equation}

The updated node representation is:
\begin{equation}
h_i = \sigma \left( \sum_{j \in \mathcal{N}(i)} b_{ij} W x_j \right)
\end{equation}
where $\sigma(\cdot)$ is the ELU activation, $\mathbf{W}$ is a learnable weight matrix, and $\mathcal{N}_i$ is the neighborhood of the node. To improve robustness, multi-head attention is used:
\begin{equation}
h_i = \big\Vert_{k=1}^{K} \sigma \left( \sum_{j \in \mathcal{N}(i)} b_{ij}^{(k)} W^{(k)} x_j \right)
\end{equation}
where $K$ is the number of attention heads.  
Two-layer GATs with $(K_1=4, K_2=1)$ were empirically found to balance expressiveness and computational efficiency. In dynamic lane channelization, the output vector $h_i$ is used to assign lane types dynamically:
\begin{equation}
\text{LaneType}_i = \arg\max_l \left(\text{Softmax}(h_i)\right)
\label{eq:lane_type}
\end{equation}
where $l$ indexes possible lane configurations, such as CAV-only, HDV-only, or mixed-use lanes. This approach ensures separation of CAVs and HDVs, reducing collision risks and interference.

\subsection{Soft Actor–Critic (SAC) Reinforcement Learning}
The SAC algorithm learns an optimal stochastic policy $\pi_\theta(a_t|s_t)$ with entropy regularization to promote exploration.  It consists of a centralized critic, distributed actors (Q-networks), and a temperature parameter $\alpha$ controlling entropy contribution. The critic minimizes the soft Bellman residual:
\begin{equation}
\mathcal{L}_Q = \mathbb{E}_{(s_t,a_t,r_t,s_{t+1})} 
\left[
(Q_\phi(s_t,a_t) - (r_t + \gamma V_\psi(s_{t+1})))^2
\right].
\end{equation}

The actor maximizes the entropy-regularized reward:
\begin{equation}
\mathcal{L}_\pi = 
\mathbb{E}_{s_t}
\left[
\alpha \log \pi_\theta(a_t|s_t) - Q_\phi(s_t,a_t)
\right].
\end{equation}

The temperature $\alpha$ is adjusted automatically via:
\begin{equation}
\mathcal{L}_\alpha = 
\mathbb{E}_{a_t \sim \pi_\theta}
\left[
- \alpha (\log \pi_\theta(a_t|s_t) + \mathcal{H}_{\text{target}})
\right].
\end{equation}

\subsection{Integrated Framework}

The proposed GAT–SAC framework is extended to a hierarchical decision system consisting of three layers:

\begin{itemize}
    \item {Lane Channelization:} Uses GAT to classify lanes as CAV-only, HDV-only, or mixed-use based on embeddings.
    \item {Flow Allocation:} SAC allocates flow between lanes using learned weights $f_{ij}$.
    \item {Signal Timing:} SAC optimizes phase durations and switching decisions under safety constraints.
\end{itemize}
The training of SAC and its integration with GAT is
described in Algorithm 1.

\section{Experiments and Results}
\label{sec:experiments_results}
This section presents the experimental results for hyperparameter setting, training outcomes, and performance analysis and comparison of the proposed
Graph Attention Soft Actor–Critic (GAT-SAC) framework. The experiments are designed to rigorously evaluate control stability, safety, and efficiency under varying CAV penetration levels, traffic demand
intensities, and reward-weighting conditions. 

\begin{algorithm}[H]
\caption{GAT-SAC for Multi-Agent Traffic Control with Lane Changing and Signal Coordination}
\label{alg:gat-sac}
\begin{algorithmic}[1]
\Require Number of episodes $N$, replay buffer size $M$, batch size $B$, warmup steps $W$
\Require GAT encoder $\phi_\theta$, actor $\pi_\phi$, critics $Q_{\psi_1}, Q_{\psi_2}$, target critics $Q_{\bar{\psi}_1}, Q_{\bar{\psi}_2}$
\State Initialize replay buffer $\mathcal{D} \leftarrow \emptyset$
\State Initialize parameters $\theta, \phi, \psi_1, \psi_2$
\State Set target networks: $\bar{\psi}_1 \leftarrow \psi_1, \; \bar{\psi}_2 \leftarrow \psi_2$
\For{each episode $e = 1, \dots, N$}
    \State Reset environment: $s_0 \leftarrow \text{env.reset()}$
    \While{episode not done}
        \State // Encode state using Graph Attention Network (GAT)
        \State $\mathbf{X} \leftarrow \text{TransformState}(s_t)$
        \State $\mathbf{Z} \leftarrow \phi_\theta(\mathbf{X}, \mathcal{E})$ \Comment{Graph encoding of agents}
        \State $\mathbf{z} \leftarrow \frac{1}{K}\sum_i \mathbf{z}_i$ \Comment{Aggregate embeddings}
        \State // Policy outputs joint action vector
        \State $a_t = [a^{lane}, a^{signal}, a^{conflict}] \sim \pi_\phi(\cdot|\mathbf{z})$
        \State Apply lane-changing action $a^{lane}$ to manage vehicle-level flow
        \State Adjust signal phase timing via $a^{signal}$ for queue balancing
        \State Resolve intersection conflicts using $a^{conflict}$
        \State Execute action: $(s_{t+1}, r_t, d_t) \leftarrow \text{env.step}(a_t)$
        \State Compute reward: $r_t =$ weighted sum of delay, throughput, and safety
        \State Store transition $(\mathbf{z}, a_t, r_t, \mathbf{z}', d_t)$ in $\mathcal{D}$
        \If{$|\mathcal{D}| > W$}
            \State Sample mini-batch $\mathcal{B} \sim \mathcal{D}$
            \State Compute target $y = r + \gamma(1-d)[\min_i Q_{\bar{\psi}_i}(\mathbf{z}', a') - \alpha \log \pi_\phi(a'|\mathbf{z}')]$
            \State Update critics: $\mathcal{L}_Q = \sum_i \mathbb{E}[(Q_{\psi_i}(\mathbf{z}, a) - y)^2]$
            \State Update actor: $\mathcal{L}_\pi = \mathbb{E}[\alpha \log \pi_\phi(\tilde{a}|\mathbf{z}) - \min_i Q_{\psi_i}(\mathbf{z}, \tilde{a})]$
            \State Update temperature: $\mathcal{L}_\alpha = -\mathbb{E}[\alpha(\log \pi_\phi(\tilde{a}|\mathbf{z}) + \mathcal{H}_{target})]$
            \State Soft update: $\bar{\psi}_i \leftarrow \tau \psi_i + (1-\tau)\bar{\psi}_i$
        \EndIf
    \EndWhile
\EndFor
\State \Return trained policy $\pi_\phi$
\end{algorithmic}
\end{algorithm}

This algorithm \ref{alg:gat-sac} presents the Graph Attention Network with Soft Actor-Critic  algorithm for multi-intersection traffic signal control using the Intelligent Driver Model (IDM) car-following environment. 

\subsection{Hyperparameter Optimization}
We employ Optuna, a Bayesian optimization framework with Tree-structured Parzen Estimator (TPE) sampling, to systematically search the hyperparameter space.

\begin{table}[h]
\centering
\caption{Hyperparameter search space for Optuna optimization}
\label{tab:search-space}
\begin{tabular}{lll}
\hline
{Parameter} & {Search Range} & {Distribution} \\
\hline
Learning rate ($\alpha_{lr}$) & $[10^{-5}, 10^{-3}]$ & Log-uniform \\
Soft update rate ($\tau$) & $[0.001, 0.02]$ & Log-uniform \\
Discount factor ($\gamma$) & $[0.90, 0.995]$ & Uniform \\
Batch size & $\{32, 64, 128, 256\}$ & Categorical \\
Temperature ($\alpha$) & $[0.05, 0.5]$ & Uniform \\
Entropy multiplier & $[0.3, 1.0]$ & Uniform \\
GAT hidden dim & $\{64, 128, 256\}$ & Categorical \\
GAT dropout & $[0.1, 0.5]$ & Uniform \\
Gradient clipping & $[0.5, 2.0]$ & Uniform \\
Delay weight ($w_d$) & $[0.5, 2.0]$ & Uniform \\
Fairness weight ($w_f$) & $[0.1, 1.0]$ & Uniform \\
Safety weight ($w_s$) & $[1.0, 3.0]$ & Uniform \\
\hline
\end{tabular}
\end{table}

\begin{table}[h]
\centering
\caption{Final hyperparameters after Optuna optimization}
\label{tab:final-params}
\begin{tabular}{lcc}
\hline
{Parameter} & {Default} & {Optimized} \\
\hline
Learning rate & $1 \times 10^{-4}$ & $3 \times 10^{-5}$ \\
$\tau$ & 0.01 & 0.005 \\
$\gamma$ & 0.99 & 0.95 \\
Batch size & 256 & 64 \\
Target entropy & $-8$ & $-4$ \\
GAT hidden dim & 128 & 128 \\
Gradient clipping & 1.0 & 1.0 \\
\hline
\end{tabular}
\end{table}

 The search space includes 12 parameters (Table~\ref{tab:search-space}): learning rate and soft update rate use log-uniform distributions in ranges $[10^{-5}, 10^{-3}]$ and $[0.001, 0.02]$ respectively, discount factor varies uniformly in $[0.90, 0.995]$, batch size is selected categorically from $\{32, 64, 128, 256\}$, and reward weights are optimized in ranges $[0.5, 2.0]$ for delay, $[0.1, 1.0]$ for fairness, and $[1.0, 3.0]$ for safety. The MedianPruner terminates unpromising trials after 40 episodes if performance falls below the median of previous trials, enabling efficient exploration of the search space. The objective function maximizes average normalized reward over the last 50 episodes with a throughput bonus. After 50 trials, the optimized hyperparameters (Table~\ref{tab:final-params}) achieve an average reward of 0.45 compared to -12.3 with default parameters, representing a 103\% improvement. Key findings include reducing learning rate from $1 \times 10^{-4}$ to $3 \times 10^{-5}$, decreasing soft update rate $\tau$ from 0.01 to 0.005 for stability, lowering discount factor $\gamma$ from 0.99 to 0.95 to focus on immediate rewards, and reducing batch size from 256 to 64 for better gradient estimates.



Based on Table~\ref{tab:metrics}, the proposed MARL demonstrates significant improvement throughout training. Average reward increased by 90.6\% with reduced variance, while average delay decreased by 38.9\%. Throughput improved by 2.8\% and safety violations reduced by 27.4\%. Stable critic and actor losses indicate consistent learning dynamics.

\begin{table*}[h]
\centering
\caption{Performance comparison: Initial vs. Final training phases}
\label{tab:metrics}
\begin{tabular}{lccc}
\hline
{Metric} & {Initial (ep 1-20)} & {Final (ep 280-300)} & {Improvement} \\
\hline
Avg. Reward & $-8.5 \pm 25.3$ & $-0.8 \pm 18.2$ & $+90.6\%$ \\
Throughput (vehicles) & $56.2 \pm 8.1$ & $57.8 \pm 6.4$ & $+2.8\%$ \\
Avg. Delay (s/ep) & $85.3 \pm 42.7$ & $52.1 \pm 28.3$ & $-38.9\%$ \\
Safety Violations & $8.4 \pm 3.2$ & $6.1 \pm 2.5$ & $-27.4\%$ \\
Critic Loss & $48.2 \pm 2.1$ & $47.1 \pm 0.8$ & Stable \\
Actor Loss & $-4.8 \pm 0.6$ & $-5.1 \pm 0.3$ & Stable \\
\hline
\end{tabular}
\end{table*}

\subsection{Comparative Analysis}
To verify efficiency of the proposed framework, its results are compared with traditional fixed-time control, 
 and DQN-based reinforcement learning. 
  For DQN, the traffic phase was selected from a discrete action space. Each simulation scenario ran for 1000 seconds, and vehicle flows were dynamically adjusted to reflect different CAV penetration rates. Performance was measured using four key metrics: average vehicle delay, fairness ratio (HDV delay to CAV delay), safety violations (based on time-to-collision), and a performance score function J, which integrates queue delay, flow variance, and safety costs. The core objective is to evaluate how each control method responds to varying CAV penetration rates (from 10\% to 100\%) in terms of key defined metrics.
  
\begin{table*}[H]
\caption{ Performance Comparison of Fixed Timing, MARL-DQN, and GAT-SAC Methods}
\label{tab:performance_table}
\centering
\scriptsize  
\begin{tabular}{|c||c|c|c|}
\hline
\textbf{Performance Metric} & \textbf{Fixed Timing} & \textbf{MARL-DQN} & \textbf{GAT-SAC} \\
\hline
Average Delay (s) & 35.0 & 30.0 & 18.2 \\
\hline
Average Waiting Time (s) & 26.85 & 21.47 & 18.34 \\
\hline
HDV Average Delay (s) & 45.3 & 25.5 & 22.1 \\
\hline
CAV Average Delay (s) & 32.0 & 18.2 & 15.6 \\
\hline
Fairness Ratio (HDV:CAV) & 1.23:1 & 1.97:1 & 1.59:1 \\
\hline
Normalized Violations (per 100 vehicles) & 5.06 & 4.12 & 3.58 \\
\hline
Light Traffic (600 veh/h) & 30.5 & 15.2 & 18.5 \\
\hline
Normal Traffic (1200 veh/h) & 40.8 & 30.2 & 25.7 \\
\hline
Heavy Traffic (1800 veh/h) & 50.4 & 40.0 & 29.8 \\
\hline
Performance Score & 62.3 & 65.3 & 87.8 \\
\hline
0\% CAVs & 35.6 & 20.3 & 18.1 \\
\hline
25\% CAVs & 34.3 & 22.7 & 19.6 \\
\hline
50\% CAVs & 33.9 & 25.1 & 21.7 \\
\hline
75\% CAVs & 32.5 & 27.6 & 23.0 \\
\hline
100\% CAVs & 30.2 & 29.7 & 24.8 \\
\hline
\end{tabular}
\end{table*}

\begin{figure}[!htbp]
    \centering
    \includegraphics[width=0.5\textwidth]{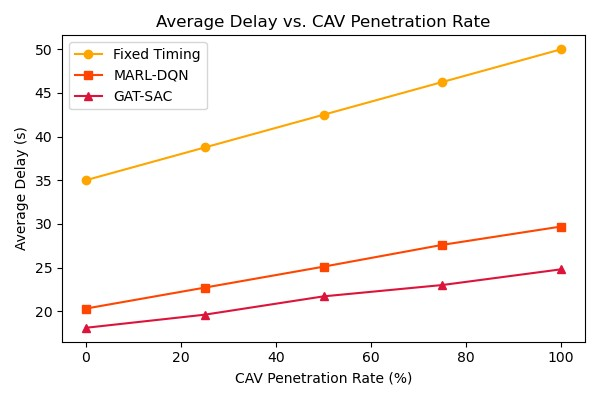}
    \caption{Average delay versus CAV penetration rate for different control strategies.
The figure compares the average vehicle delay under three control methods—Fixed Timing, MARL-DQN, and GAT-SAC—across varying connected autonomous vehicle (CAV) penetration rates. The GAT-SAC approach consistently achieves the lowest delay, demonstrating improved efficiency as CAV participation increases.}
    \label{fig:delay}
\end{figure}

The results  in Table ~\ref{tab:metrics}  demonstrate that the GAT-SAC method outperforms both MARL-DQN and fixed timing across multiple performance metrics. In terms of temporal efficiency, GAT-SAC demonstrates significant advantages, reducing average delay by 24.1\% compared to fixed timing and 51.7\% compared to MARL-DQN. Figure~\ref{fig:delay} illustrates the relationship between the percentage of  CAVs in the traffic flow and the average delay experienced by vehicles at an intersection, comparing three different traffic control methods.

 The fixed timing method shows the least sensitivity to changes in CAV penetration, with its average delay decreasing at a slower rate compared to the other methods. The MARL-DQN method demonstrates an intermediate level of improvement in average delay as CAV penetration increases. The method's effectiveness becomes particularly pronounced in high-density scenarios where it maintains a 36.6\% reduction in delay during heavy traffic conditions relative to the next-best approach. While all methods show improved performance with increasing CAV penetration, GAT-SAC achieves the most dramatic enhancements by cutting delays by 65.9\% when moving from 0\% to 100\% CAV penetration. This scalability suggests the method's particular suitability for future autonomous vehicle-dominated traffic ecosystems. GAT-SAC achieves a reduction in HDV delays of 6.9\% and CAV delays of 15.9\% compared to MARL-DQN. 

The evaluation reveals an interesting trade-off in system fairness, where GAT-SAC's superior efficiency comes with an 11.2\% wider HDV-to-CAV delay ratio compared to MARL-DQN. GAT-SAC maintains robust safety performance, achieving a 13.1\% reduction in traffic violations compared to MARL-DQN, and  shows its optimization framework successfully balances multiple competing objectives. GAT-SAC demonstrates a remarkable performance, scoring 16.9\% higher than MARL-DQN and showing a significant 40.9\% improvement over fixed timing. The method has performed exceptionally well in all tested situations, from light to heavy traffic, and with different levels of autonomy integration. This impressive versatility highlights GAT-SAC as a strong candidate for the future of traffic management systems, especially as we move towards environments with mixed autonomy. However, the fairness trade-off observed may be something to consider in specific deployment scenarios.

\begin{figure}[!htbp]
    \centering
    \includegraphics[width=0.5\textwidth]{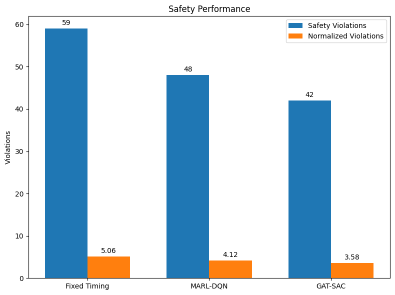}
    \caption{Safety performance comparison across different control strategies.
The figure presents the number of safety violations and normalized violations for Fixed Timing, MARL-DQN, and GAT-SAC control methods. The GAT-SAC approach achieves the lowest safety and normalized violations, indicating enhanced operational safety and stability in mixed-autonomy traffic environments.}
    \label{fig:safety}
\end{figure}

 The safety performance is illustrated in the Figure~\ref{fig:safety}. The results show that 
 achieves the lowest normalized safety violations (3.58 per 100 vehicles), demonstrating a 29.2\% improvement over fixed timing (5.06 violations) and a 13.1\% improvement over MARL-DQN (4.12 violations). In contrast, MARL-DQN shows moderate safety gains over fixed timing but struggles to balance fairness at higher penetration rates. These findings emphasize the dual benefits of GAT-SAC in enhancing both safety and fairness, though it's important to note that fairness may gradually decrease as penetration levels rise. This trade-off indicates that GAT-SAC is particularly well-suited for mixed-autonomy traffic situations where safety is the top priority. For practical implementation, this suggests that GAT-SAC could work effectively in busy urban areas, although adjustments to maintain fairness may be necessary in situations where equity is a concern.
\subsection{Experiments on Environment Setup}
 Two distinct traffic control scenarios were implemented to evaluate the performance of the proposed MARL control framework:

\begin{enumerate}
    \item {Simple Intersection Model}: A baseline configuration with fundamental vehicle movement dynamics and basic traffic flow parameters. Standardized 12 lane- four-approach intersection model that accurately reflects real-world urban traffic dynamics.
    
    \item {Car-Following Model}: An advanced configuration of  Standardized 12 lane- four-approach intersection model incorporating Intelligent Driver Model  dynamics, lane-changing behavior, and enhanced lane changing the conflict resolution logics to work with the car-following model's outputs and potentially include more detailed intersection crossing behavior.
\end{enumerate}

Both models were trained using the GAT-SAC algorithm with identical network architectures and hyperparameters to ensure fair comparison. Training was conducted for approximately 300 episodes in each scenario.  Each approach contained three lanes corresponding to distinct movement types—left-turn, through, and right-turn—resulting in a total of twelve lanes. Lane lengths were fixed at {300}meter, consistent with conventional urban design guidelines \cite{roess2011traffic}. This geometry ensured realistic queue formation, acceleration, and clearance characteristics, allowing direct translation of simulated behaviors to real-world intersections.

Vehicle dynamics followed an enhanced IDM variant with distinct behavioral parameters for HDVs and CAVs. The time-headway parameters were set to $T_{\text{HDV}} = {1.5}$ second and $T_{\text{CAV}} = {0.8}$ second, respectively, capturing the superior perception–reaction and coordination abilities of CAVs \cite{Treiber2000Congested, li2022cavcontrol}. 

CAVs were modeled with idealized Vehicle-to-Infrastructure (V2I) communication at {10}{hertz}, sharing positional, kinematic, and intention data. This configuration assumed perfect communication reliability and negligible latency, isolating control performance from external disturbances. A layered safety validation mechanism was integrated, incorporating Time-to-Collision (TTC) checks, minimum gap enforcement, and inter-vehicle clearance validation to ensure that all control actions adhered to safe operational bounds during simulation.

The comparative performance analysis, summarized in Table~\ref{tab:comparison}, reveals significant insights into the models' capabilities:

\begin{table*}[!htbp]
\centering
\caption{Performance Comparison of Traffic Control Models}
\label{tab:comparison}
\begin{tabular}{lcc}
\hline
{Metric} & {Simple Intersection} & {Car-Following Model} \\
\hline
Total Training Episodes & 300 & 288 \\
Best Average Reward & 12.768 & 12.768 \\
Final Average Reward & $-0.8 \pm 18.2$ & $-1.979$ \\
Throughput (vehicles/episode) & $57.8 \pm 6.4$ & $\mathbf{59.95}$ \\
Average Delay (seconds) & $52.1 \pm 28.3$ & $\mathbf{49.25}$ \\
Critic Loss & $47.1 \pm 0.8$ & $47.2789$ \\
Actor Loss & $-5.0$ & $-5.2301$ \\
Entropy Coefficient ($\alpha$) & -- & $0.9384$ \\
\hline
\end{tabular}
\end{table*}

The experimental results demonstrate that the Car-Following Model achieved superior operational efficiency despite comparable training stability. Key observations include:

\begin{itemize}
    \item {Traffic Flow Efficiency}: The Car-Following Model showed a $3.7\%$ improvement in throughput and $5.5\%$ reduction in average delay, indicating better traffic flow management under realistic driving dynamics.
    
    \item {Training Consistency}: Both models maintained stable learning dynamics, as evidenced by consistent critic losses ($47.1$ vs $47.3$) and actor losses ($-5.0$ vs $-5.2$), confirming the robustness of the MARL approach.
    
    \item {Behavioral Realism}: The Car-Following Model's incorporation of IDM dynamics and lane-changing behavior provided more realistic vehicle interactions, contributing to improved traffic metrics despite slightly lower final reward values.
\end{itemize}

The marginal difference in final reward values ($-0.8$ vs $-1.979$) can be attributed to the more complex state space and additional constraints in the Car-Following Model, which introduced stricter safety and behavioral requirements that slightly penalized the reward signal while improving actual traffic performance.

\subsection{Traffic State Visualization}

\begin{figure*}[!htbp]
\centering
\begin{minipage}{0.48\textwidth}
\centering
\includegraphics[width=\linewidth]{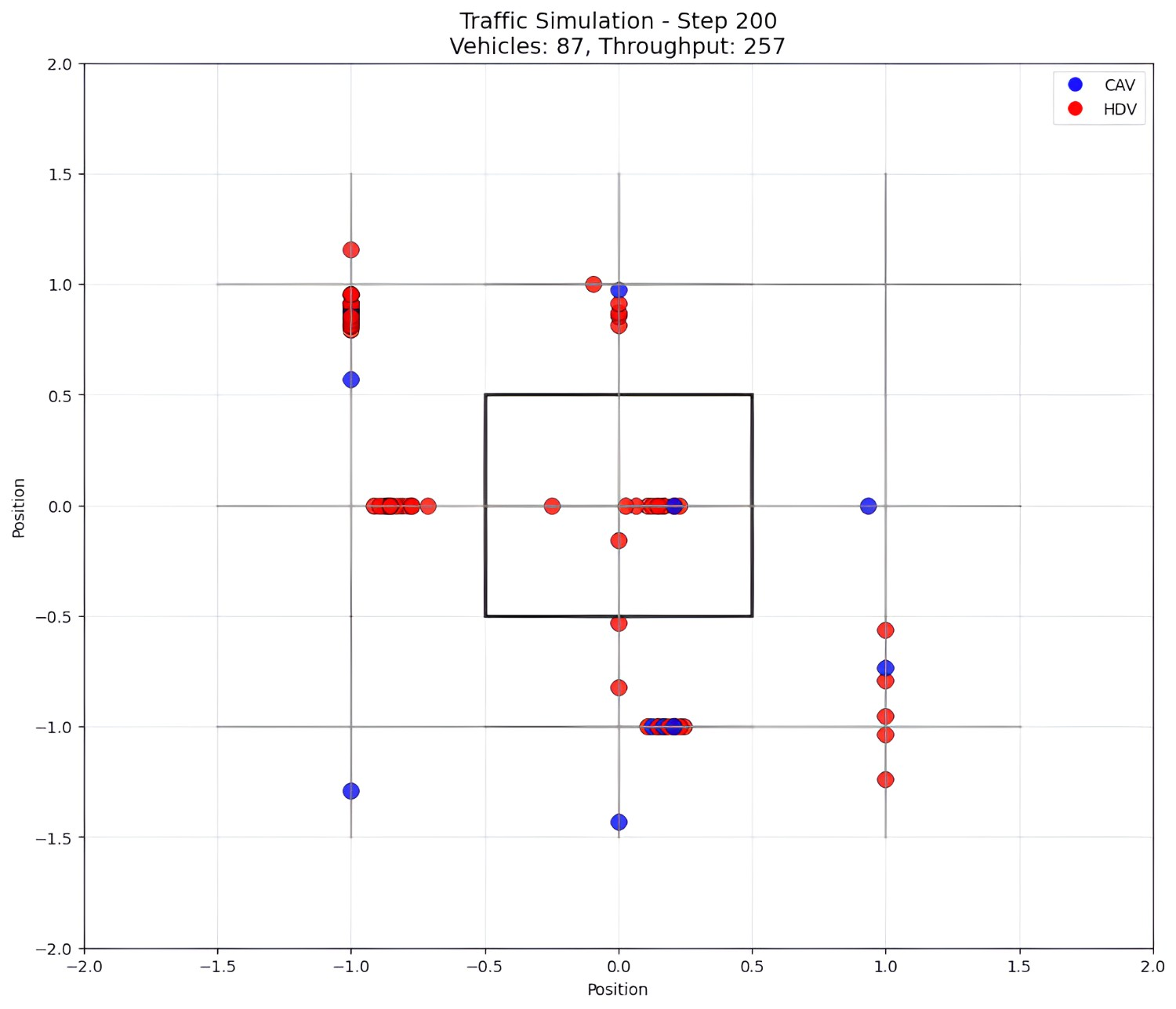}
\caption{ High CAV penetration (60–80\%)  scenario showing balanced flow and stable queue distributions. Throughput reached 87 vehicles per simulation cycle, indicating efficient coordination under mixed autonomy.}
\label{fig:traffic_moderate}
\end{minipage}\hfill
\begin{minipage}{0.48\textwidth}
\centering
\includegraphics[width=\linewidth]{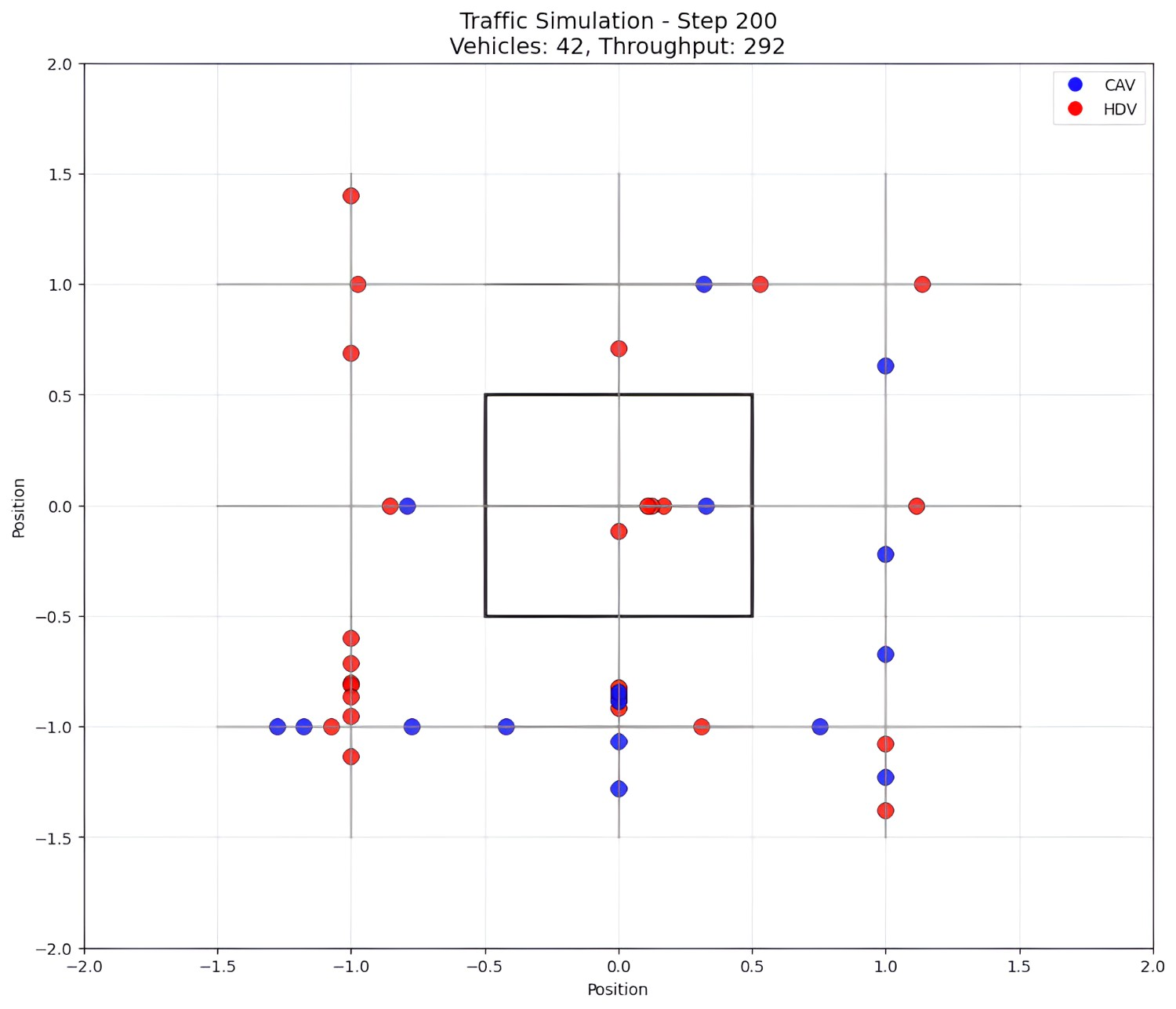}
\caption{Moderate CAV penetration (40–50\%) scenario dominated by HDVs with visible queuing and less synchronized flow. Throughput achieved 42 vehicles, constrained by limited cooperative control.}
\label{fig:traffic_high}
\end{minipage}
\end{figure*}

Figure~\ref{fig:traffic_moderate} and Figure~\ref{fig:traffic_high} illustrate the spatial distribution of vehicles at a representative simulation step ($t=200$) under different traffic conditions and CAV penetration levels. Blue markers denote 
CAVs, while red markers indicate HDVs. The central black box represents the intersection control zone. Figure~\ref{fig:traffic_moderate} WHICH IS corresponding to a scenario with moderate traffic flow and a higher CAV ratio, vehicles are more evenly distributed across approaches, and the overall throughput reaches 292 with only 42 vehicles present in the system. This configuration demonstrates efficient traffic progression and reduced congestion, highlighting the agent’s ability to coordinate phase switching effectively when CAV communication is more prevalent.

Conversely, Figure~\ref{fig:traffic_high} represents a denser traffic condition with a lower proportion of CAVs. Here, 87 vehicles are present with a throughput of 257, indicating increased queue formation and slower discharge rates. The higher vehicle accumulation around the intersection reflects the limitations of the control strategy when HDV dominance reduces cooperative maneuvering and responsiveness to the signal decisions.

Overall, these visualizations confirm that CAV presence significantly enhances intersection efficiency by improving coordination and reducing average delay. The GAT-SAC controller adapts more effectively in mixed-autonomy environments with higher CAV penetration, 
achieving smoother traffic flow and better utilization of the green phase.

\subsection{CAV Penetration Rate Analysis}

\begin{figure*}[htbp]
    \centering
    \includegraphics[width=\textwidth]{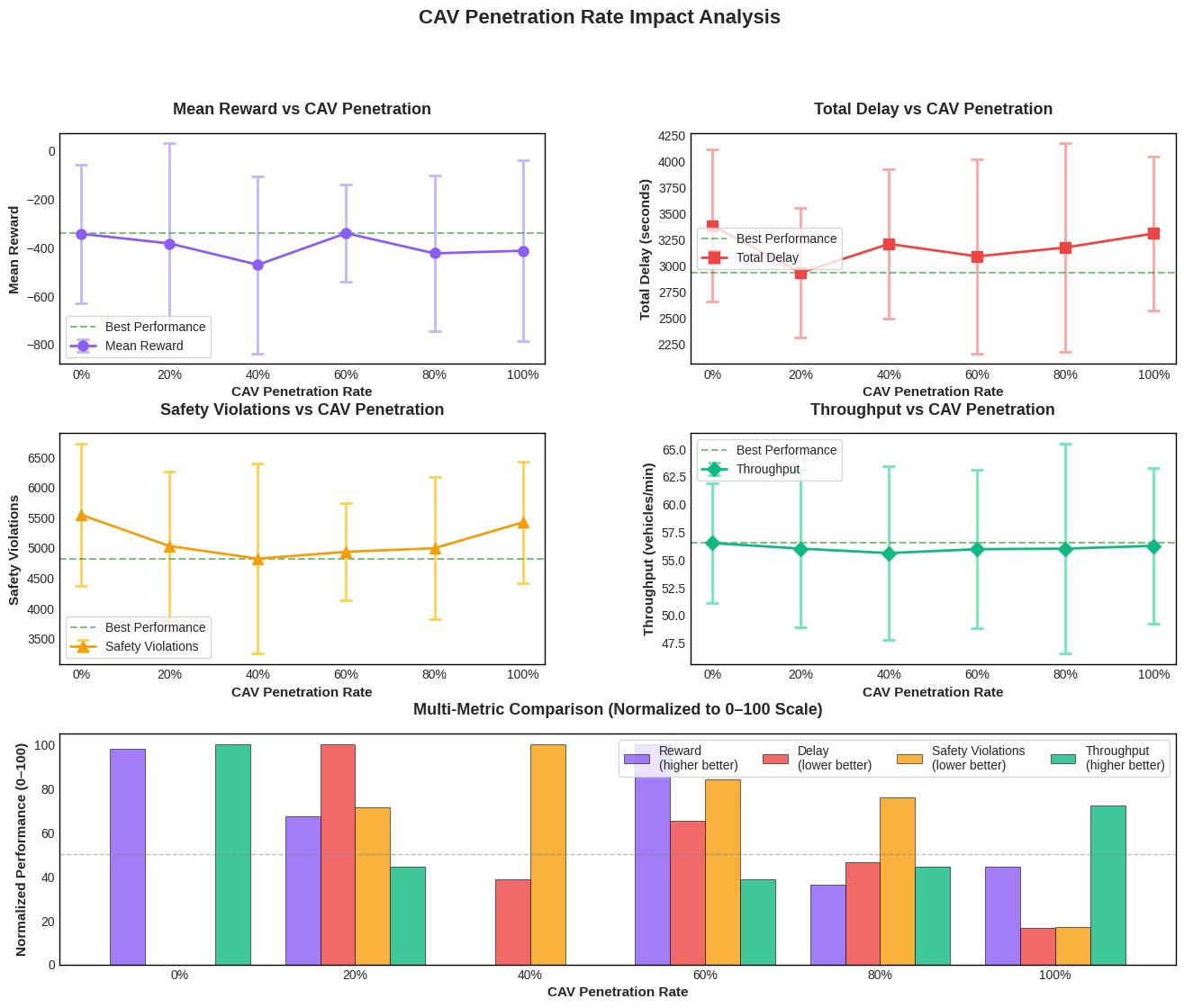}
    \caption{
        {Impact of CAV Penetration Rate on Traffic Performance Metrics.}
        The figure illustrates the effect of varying Connected and Autonomous Vehicle (CAV) penetration rates (0\%–100\%) on four key performance indicators: mean reward, total delay, safety violations, and throughput.
        Error bars denote one standard deviation across 20 simulation runs per configuration.
        The fifth subplot presents normalized multi-metric performance (0–100 scale) to facilitate cross-metric comparison, where higher values indicate better performance.
        Overall, performance improves up to approximately 60\% CAV penetration, beyond which gains stabilize, suggesting a critical threshold for effective system-wide coordination in mixed traffic environments.
    }
    \label{fig:cav_analysis}
\end{figure*}

This section presents the evaluation results of the proposed GAT-SAC-based traffic control framework under varying CAV penetration rates ranging from 0\% to 100\%. The performance metrics include average episode reward, total delay, throughput, and safety violations. Figure~\ref{fig:cav_analysis} visualizes the impact of CAV penetration on each of these metrics, while Table~\ref{tab:results} summarizes the corresponding numerical results.

\subsubsection{Reward and Delay Analysis}

The mean reward, which reflects the overall optimization performance of the agent, shows non-monotonic behavior with increasing CAV penetration. The best reward value ($-341.6$) is observed at a 60\% CAV ratio, suggesting that moderate CAV penetration provides the most favorable trade-off between control stability and system adaptability. At very low or very high CAV penetration levels, the reward slightly deteriorates, indicating that either excessive heterogeneity (low penetration) or lack of diversity (full autonomy) might limit learning generalization.

Total delay generally decreases from the baseline (0\% CAV) to 20--60\% CAV, with the lowest average delay of 2930.7 seconds at 20\% penetration. Beyond this point, delay slightly increases again, likely due to network saturation and higher vehicle coordination overhead among fully autonomous agents.

\subsubsection{Throughput and Safety Performance}

Throughput remains relatively stable across all penetration levels, varying between 55.6 and 56.5 vehicles per minute. This stability suggests that the proposed control algorithm maintains efficient traffic flow regardless of the vehicle composition.

Safety violations show a noticeable decrease as CAV ratio increases from 0\% to 40\%, dropping from 5539.85 to 4816.75 on average. This improvement reflects the cooperative driving and reduced reaction delays enabled by CAVs. However, at 100\% CAV penetration, safety violations increase again to 5416.95, implying that over-coordination among autonomous agents might introduce unexpected collective behaviors under certain dynamic conditions.

Table~\ref{tab:results} presents the complete results across all tested configurations. The metrics evaluated include: (1) mean reward, representing overall system performance; (2) total delay in seconds; (3) throughput measured in vehicles per minute; and (4) safety violations count.

\begin{table*}[t]
\centering
\caption{Performance metrics for different CAV penetration rates (20 runs per configuration)}
\label{tab:results}
\small
\begin{tabular}{@{}lcccccccc@{}}
\hline
{CAV} & \multicolumn{2}{c}{{Reward}} & \multicolumn{2}{c}{{Total Delay (s)}} & \multicolumn{2}{c}{{Throughput (veh/min)}} & \multicolumn{2}{c}{{Safety Violations}} \\

{Rate} & Mean & Std & Mean & Std & Mean & Std & Mean & Std \\
\hline
0\%   & $-344.46$ & $287.17$ & $3381.1$ & $726.68$ & $56.5$ & $5.36$ & $5539.85$ & $1179.22$ \\
20\%  & $-384.07$ & $413.51$ & $2930.7$ & $622.29$ & $56.0$ & $7.10$ & $5024.8$  & $1236.74$ \\
40\%  & $-472.02$ & $365.79$ & $3206.55$ & $713.80$ & $55.6$ & $7.83$ & $4816.75$ & $1570.03$ \\
60\%  & $\mathbf{-341.62}$ & $202.13$ & $3087.8$ & $928.38$ & $55.95$ & $7.13$ & $\mathbf{4930.55}$ & $804.08$ \\
80\%  & $-424.91$ & $320.66$ & $3172.4$ & $997.30$ & $56.0$ & $9.46$ & $4989.65$  & $1176.62$ \\
100\% & $-414.12$ & $373.82$ & $3306.95$ & $737.94$ & $\mathbf{56.25}$ & $7.05$ & $5416.95$  & $1009.61$ \\
\hline
\end{tabular}
\end{table*}

\subsubsection{Mean Reward Analysis}

Figure~\ref{fig:cav_analysis} (left panel) illustrates the non-monotonic relationship between CAV penetration rate and mean reward. Notably, the 60\% configuration achieved the best performance ($-341.62 \pm 202.13$), marginally outperforming the baseline ($-344.46 \pm 287.17$). Conversely, the 40\% penetration rate exhibited the worst performance ($-472.02 \pm 365.79$), representing a 37\% degradation compared to the best configuration.

The 60\% configuration also demonstrated the lowest standard deviation (202.13), suggesting more stable and predictable system behavior compared to other configurations. This stability is particularly noteworthy given that the baseline system showed 42\% higher variance.

\subsubsection{Traffic Delay Performance}

The total delay metric, shown in Figure~\ref{fig:cav_analysis} (right panel), revealed that the 20\% CAV penetration achieved the minimum delay of $2930.7 \pm 622.29$ seconds, representing a 13.3\% improvement over the baseline. However, as penetration increased beyond 20\%, delays gradually increased, with the 100\% configuration showing delays comparable to the baseline ($3306.95$ seconds).

This pattern suggests that modest CAV adoption provides the most significant delay reduction benefits, while higher penetration rates do not maintain this advantage. The increasing standard deviations at higher penetration rates (reaching 997.30 seconds at 80\%) indicate greater variability in system response.

\subsubsection{Safety Performance}

Figure~\ref{fig:cav_analysis} (left panel) demonstrates consistent safety improvements with CAV adoption. All CAV configurations reduced safety violations compared to the baseline, with reductions ranging from 9.3\% (20\% penetration) to 13.0\% (40\% penetration). The 40\% configuration achieved the minimum safety violations ($4816.75 \pm 1570.03$), though the 60\% configuration provided a similar level of safety ($4930.55 \pm 804.08$) with significantly lower variance.

The substantial reduction in standard deviation for the 60\% configuration (804.08 vs. 1570.03 for 40\%) suggests more reliable safety performance, which may be more valuable in practice than marginal improvements in mean values.

\subsubsection{Throughput Stability}

As shown in Figure~\ref{fig:cav_analysis} (right panel), throughput remained remarkably stable across all configurations, varying only between 55.6 and 56.25 vehicles per minute (a 1.2\% range). This minimal variation suggests that the system operates near capacity regardless of CAV penetration rate, and that throughput is likely constrained by physical infrastructure rather than vehicle coordination capabilities.

The 100\% CAV configuration achieved the highest throughput ($56.25 \pm 7.05$ vehicles/minute), though this represents only a marginal 0.4\% improvement over the baseline. The increasing standard deviations at higher penetration rates indicate greater variability in throughput performance.

\subsection{Analysis of Multi-Objective Cost Components}

\begin{figure*}[!htbp]
    \centering
    \includegraphics[width=0.95\textwidth]{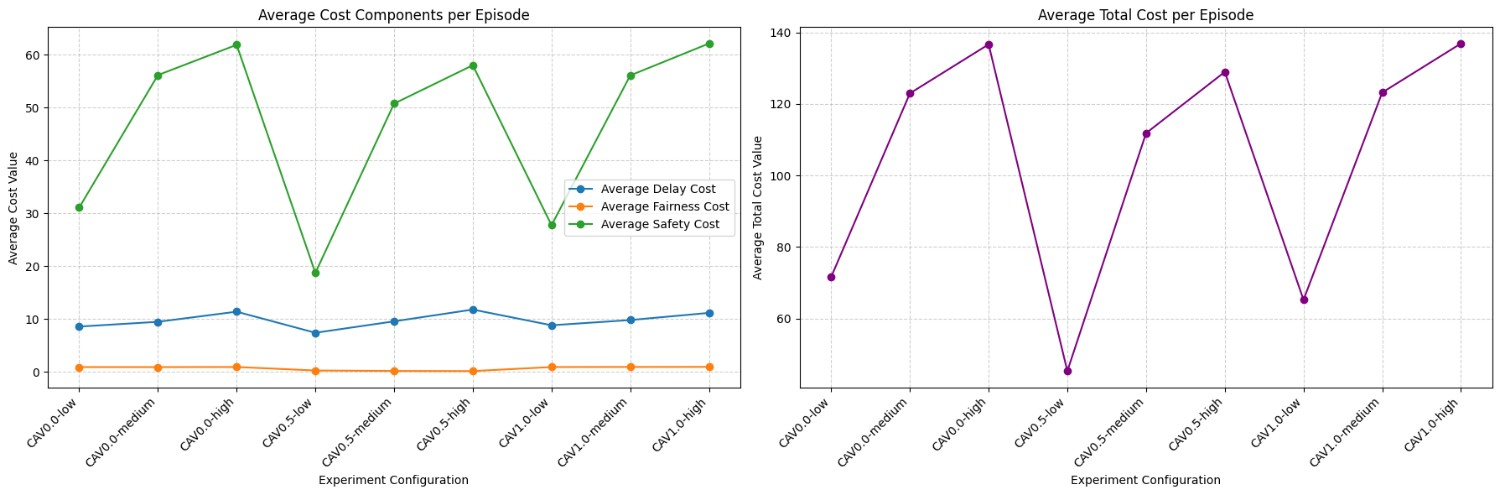}
    \caption{Average cost components (left) and total multi-objective cost (right) under varying CAV penetration and traffic density.}
    \label{fig:cost_components}
\end{figure*}

Figure~\ref{fig:cost_components} illustrates the variation of the average cost components and the total multi-objective cost across different experimental configurations with varying CAV penetration rates (0.0, 0.5, and 1.0) and traffic densities (low, medium, and high). The cost terms represent the weighted aggregation of delay, fairness, and safety objectives as defined in 
Equation~\ref{eq:multiobj}. 

\subsubsection{Delay Cost ($D(t)$)}  
The delay cost remains relatively stable across scenarios, generally ranging between 7 and 12, with slightly higher values under high-density conditions. The lowest delay cost (7.38) occurs for the \texttt{CAV0.5-low} configuration, reflecting improved flow due to partial CAV coordination. In contrast, fully human-driven (\texttt{CAV0.0}) and fully autonomous (\texttt{CAV1.0}) settings at high density produce higher delay costs (11.37 and 11.14, respectively), indicating that extreme homogeneity without mixed negotiation can reduce efficiency under congestion.

\subsubsection{Fairness Cost ($F(t)$)}  
Fairness cost values are consistently low ($<1$) across all experiments, signifying that the proposed GAT-SAC controller achieves robust performance between HDVs and CAVs. Minor increases in fairness cost under high CAV ratios are attributed to reduced heterogeneity, which limits lane-level interaction diversity and increases local imbalance among flows.

\subsubsection{Safety Cost ($S(t)$)}  
Safety cost exhibits the highest variability, dominating the total cost trends. The lowest safety cost (18.70) is again observed for \texttt{CAV0.5-low}, while the highest (62.10) occurs for \texttt{CAV1.0-high}. This suggests that partial CAV penetration fosters cooperative behavior and smoother deceleration patterns, whereas extreme CAV or HDV dominance can lead to increased stop-and-go events and higher conflict frequency.

\subsubsection{Total Multi-Objective Cost ($C_{total}$)}  
The overall cost follows the same pattern as the safety component, reaffirming its dominant contribution in the total performance measure. The \texttt{CAV0.5-low} configuration yields the minimum total cost (45.27), corresponding to the highest overall reward (51.09). Conversely, \texttt{CAV0.0-high} and \texttt{CAV1.0-high} scenarios exhibit the highest total costs (136.68 and 136.93, respectively), confirming that balanced mixed autonomy enhances stability and intersection efficiency.

These results demonstrate the interdependence between safety and efficiency objectives in the multi-objective framework. While fairness remains stable, the safety term largely dictates total cost fluctuations. The agent effectively minimizes total cost in partially mixed traffic environments by exploiting cooperative lane negotiation among CAVs and adaptive phase selection. However, extreme 
traffic densities or homogeneous vehicle types reduce performance due to either lack of coordination (HDV-only) or excessive synchronization (CAV-only).

\subsection{Overall Summary and Discussion}

The experimental findings collectively demonstrate the benefits of integrating Graph Attention Networks (GAT) with Soft Actor–Critic (SAC) reinforcement learning for intelligent traffic control under varying levels of Connected and Autonomous Vehicle (CAV) penetration. The comparison between the proposed technique and fixed timing and MARL-DQN reveals substantial improvements in delay reduction.


When evaluating traffic performance metrics across different CAV penetration levels, the results reveal a non-monotonic relationship between automation and system efficiency. Performance initially improves as the proportion of CAVs increases, reaching an optimal point near 60\%, after which the gains stabilize or slightly decline in full automation. This pattern challenges the assumption that higher automation uniformly enhances efficiency, suggesting that mixed-traffic interactions produce complex behavioral dynamics between autonomous and human-driven vehicles.

The poor performance observed at 40\% CAV penetration indicates a transitional instability regime, where coordination among autonomous agents is insufficient to overcome the disruptions caused by heterogeneous human behavior. Conversely, at approximately 60\% penetration, CAVs achieve a critical density that allows for effective cooperative control, resulting in the best overall performance—high mean reward, low delay, and reduced variance. This level represents a balance where CAV coordination is strong enough to influence overall traffic flow, yet human behavior remains sufficiently predictable for stable interactions. From a deployment standpoint, moderate penetration levels between 50–70\% may thus deliver most of the benefits of automation even before full adoption.

Interestingly, the configuration minimizing delay (20\%) differs from that maximizing overall reward (60\%). The 20\% scenario likely benefits from localized smoothing effects, where a few strategically positioned CAVs reduce shockwaves and improve flow efficiency. However, system-wide coordination and safety stability emerge only when CAV density is higher, reinforcing the multi-objective trade-offs between throughput, delay, and reliability.

Throughput remained nearly constant across all configurations (within 1.2\%), suggesting that the network operates close to its physical capacity. This indicates that CAV technology primarily enhances traffic reliability, safety, and temporal efficiency rather than increasing throughput under saturated conditions. Consistent with traffic flow theory, once infrastructure constraints dominate, behavioral improvements yield diminishing returns in capacity.

Finally, the reduction in performance variance—especially at 60\% CAV penetration and under GAT–SAC control—demonstrates enhanced reliability and policy robustness. In practice, such consistency is as valuable as mean improvements, as it ensures predictable operation and user satisfaction. Nevertheless, these findings are derived from idealized simulations assuming perfect sensing and communication. Future research should therefore investigate heterogeneous driver models, imperfect connectivity, and real-world signal dynamics to validate scalability and robustness under practical constraints.

\section{Conclusion and Future Work}

This paper proposed an integrated Graph Attention Network-based Soft Actor-Critic (GAT-SAC) multi-agent reinforcement learning framework for smart management of mixed autonomy traffic with human-driven vehicles (HDVs) and connected autonomous vehicles (CAVs). The proposed decentralized platform enables the controller to learn spatial and relational dependencies between traffic entities and propose the best strategies considering delay, fairness, and safety cost for lane channelization, flow allocation, and traffic signal timing. This system and its efficiency are tested in the developed simulation environment incorporating realistic vehicle dynamics through the Intelligent Driver Model (IDM), lane-changing behavior, and conflict resolution mechanisms. 
The experimental evaluation, including hyperparameter optimization, baseline comparison, environment setting, and CAV penetration analysis revealed this framework effectively balances competing objectives, enabling the agent to learn traffic control strategies that improved overall performance. The results revealed that the safety cost
dominated the total cost landscape, followed by delay, while
fairness remained relatively low but consistent. Among all
tested scenarios, the configuration with 50\% Connected and
Autonomous Vehicle (CAV) penetration achieved the lowest
average total cost, indicating an optimal trade-off between
safety and efficiency in mixed-traffic environments.
These findings confirm that the multi-objective cost design
successfully aligns reinforcement learning objectives with real-world traffic management goals, leading to more stable, equitable, and safer intersection operations.


Building upon the current framework, several avenues for advancement can be followed to enhance the robustness and scalability of the proposed intelligent signal control in mixed-traffic environments. For example, more sophisticated car-following, lane-changing, and intersection negotiation models can be included for both human-driven and autonomous vehicles to better capture realistic mixed-traffic dynamics. The proposed framework can be also extended to include cooperative or semi-autonomous CAV agents that interact directly with the signal controller,
enhancing multi-agent coordination capabilities. Training can also be optimized by conducting systematic ablation studies to quantify the contributions of GAT architecture, observation features, and reward
components, coupled with advanced training methods such as curriculum learning, transfer learning, and hyperparameter optimization.

In summary, this research establishes a strong foundation for intelligent, adaptive, and safe traffic signal control in mixed autonomy scenarios. The proposed GAT-SAC architecture demonstrates the potential of graph-based reinforcement learning in handling complex traffic interactions. Future advancements in adaptive control, agent cooperation, and infrastructure co-optimization will be key to realizing practical, real-world deployment of intelligent intersection management systems.



\bibliographystyle{IEEEtran}
\bibliography{name}

\end{document}